\documentclass[final,5p,times,twocolumn]{elsarticle}

\usepackage{hyperref}
\hypersetup{colorlinks=true}
\usepackage{amssymb}
\usepackage{amsmath}
\usepackage{amsthm}
\usepackage{graphicx}
\usepackage{caption}
\usepackage{subcaption}
\journal{Fusion Engineering and Design}

\begin{document}

\begin{frontmatter}

%% Title, authors and addresses

%% use the tnoteref command within \title for footnotes;
%% use the tnotetext command for the associated footnote;
%% use the fnref command within \author or \address for footnotes;
%% use the fntext command for the associated footnote;
%% use the corref command within \author for corresponding author footnotes;
%% use the cortext command for the associated footnote;
%% use the ead command for the email address,
%% and the form \ead[url] for the home page:
%%
%% \title{Title\tnoteref{label1}}
%% \tnotetext[label1]{}
%% \author{Name\corref{cor1}\fnref{label2}}
%% \ead{email address}
%% \ead[url]{home page}
%% \fntext[label2]{}
%% \cortext[cor1]{}
%% \address{Address\fnref{label3}}
%% \fntext[label3]{}

\title{Impact of nuclear irradiation on helium bubble nucleation at interfaces in liquid metals coupled to permeation through stainless steels}

%% use optional labels to link authors explicitly to addresses:
%% \author[label1,label2]{<author name>}
%% \address[label1]{<address>}
%% \address[label2]{<address>}

\author[1]{J. Fradera\corref{cor1}\fnref{labelc1}}
\author[1]{S.Cuesta-L\'{o}pez\fnref{labelc2}}
\address[1]{Advanced Materials, Nuclear Technology, Applied Nanotechnology, University of Burgos (UBU), Science and Technology Park, I+D+I Building, Room 63, Plaza Misael Bañuelos, s/n, 09001, Burgos, Spain}
\cortext[cor1]{corresponding author}
\fntext[labelc1]{contact: jfradera@ubu.es}
\fntext[labelc2]{correspondence may also be sent to: scuesta@ubu.es}
\newcommand{\OF}{OpenFOAM\textsuperscript{\textregistered} }

\begin{abstract}

The impact of nucleating gas bubbles in the form of a dispersed gas phase on hydrogen isotope permeation at interfaces between liquid metals, like LLE , and structural materials, like stainless steel, has been studied. Liquid metal to structural material interfaces involving surfaces, may lower the nucleation barrier promoting bubble nucleation at active sites. Hence, hydrogen isotope absorption into gas bubbles modelling and control at interfaces may have a capital importance regarding design, operation and safety.

He bubbles as a permeation barrier principle is analysed showing a significant impact on hydrogen isotope permeation, which may have a significant effect on liquid metal systems, e.g., tritium extraction systems. Liquid metals like LLE under nuclear irradiation in, e.g., breeding blankets of a nuclear fusion reactor would generate tritium which is to be extracted and recirculated as fuel. At the same time that tritium is bred, helium is also generated and may precipitate in the form of nano bubbles. 

Phenomena modelling is exposed and implemented in \OF CFD tool for 0D to 3D simulations. Results for a 1D case show the impact of a He dispersed phase of nano bubbles on hydrogen isotopes permeation at an interface. In addition, a simple permeator simulation, consisting in a straight 3D pipe is exposed showing the effect of a He dispersed gas phase on hydrogen isotope permeation through different stainless steels. Results show the permeation reduction as a function of the interface area covered by He bubbles.

Our work highlights the effect of gas bubble nucleation at interfaces and the importance of controlling these phenomena in nuclear technology applications.

\end{abstract}

\begin{keyword}
%% keywords here, in the form: keyword \sep keyword

%% MSC codes here, in the form: \MSC code \sep code
%% or \MSC[2008] code \sep code (2000 is the default)

\end{keyword}

\end{frontmatter}
\newcommand{\OF}{OpenFOAM\textsuperscript{\textregistered} }

%%
%% Start line numbering here if you want
%%
% \linenumbers

%% main text
\section*{Glossary}
\newcommand{\Item}[2]{\item[\textbf{#1\hfill}] #2}
\newcommand{\Itema}[2]{\item[\hfill #1] #2}

\subsection*{Abbreviations}
\begin{list}{}{%
\settowidth{\labelwidth}{\textbf{$A,B,C,D$}}%
\setlength{\labelsep}{2. em}%
\setlength{\leftmargin}{\labelwidth}%
\addtolength{\leftmargin}{\labelsep}%
\setlength{\rightmargin}{0. cm}%
\setlength{\parsep}{\parskip}%
\setlength{\itemsep}{0. cm}\setlength{\topsep}{0. cm}%
\setlength{\partopsep}{0. cm}}

\Item{CFD}{Computational Fluid Dynamics}
\Item{CC}{Cooling Channels}
\Item{CP}{Cooling Plates}
\Item{EoS}{Equation of State}
\Item{EU'97}{EUROFER'97}
\Item{FEM}{Finite Element Method}
\Item{FVM}{Finite Volume Method}
\Item{HCLL}{Helium--Cooled Lithium Lead}
\Item{LLE}{Lithium Lead Eutectic alloy}
\Item{LM}{Liquid Metal}
\Item{RAFM}{Reduced Activation Ferritic Martensitic}
\Item{SM}{Structural Material}
\Item{SP}{Stiffening Plates}
\Item{T}{Tritium}
\end{list}

\subsection*{Greek characters}
\begin{list}{}{%
\settowidth{\labelwidth}{\textbf{$A,B,C,D$}}%
\setlength{\labelsep}{2. em}%
\setlength{\leftmargin}{\labelwidth}%
\addtolength{\leftmargin}{\labelsep}%
\setlength{\rightmargin}{0. cm}%
\setlength{\parsep}{\parskip}%
\setlength{\itemsep}{0. cm}\setlength{\topsep}{0. cm}%
\setlength{\partopsep}{0. cm}}

\Item{$\theta$}{contact angle}
\Item{$\kappa$}{effective solubility ratio}
\Item{$\pi$}{number pi}
\Item{$\rho$}{Density}
\Item{$\sigma$}{Surface tension}
\Item{$\upsilon_0$}{volume of one atom or molecule}
\Item{$\psi$}{supersaturation ratio}

\end{list}

\subsection*{Latin characters}
\begin{list}{}{%
\settowidth{\labelwidth}{\textbf{$A,B,C,D$}}%
\setlength{\labelsep}{2. em}%
\setlength{\leftmargin}{\labelwidth}%
\addtolength{\leftmargin}{\labelsep}%
\setlength{\rightmargin}{0. cm}%
\setlength{\parsep}{\parskip}%
\setlength{\itemsep}{0. cm}\setlength{\topsep}{0. cm}%
\setlength{\partopsep}{0. cm}}

\Item{$a$}{specific area}
\Item{$f(\theta)$}{shape factor}
\Item{$\Delta g$}{nucleation driving force}
\Item{$k_{B}$}{Boltzmann's constant}
\Item{$k_{r}$}{recombination coefficient}
\Item{$k_{S}$}{Sievert's coefficient}
\Item{$m_0$}{mass of one atom or molecule}
\Item{$n$}{material depending exponent}
\Item{$r$}{radial coordinate}
\Item{$r_b$}{bubble radius}
\Item{$t$}{time}
\Item{$\textbf{u}$}{fluid velocity}
\Item{$C$}{concentration}
\Item{$D$}{diffusivity}
\Item{$G$}{Gibbs free energy}
\Item{$M$}{molar mass}
\Item{$N_b$}{concentration of bubbles}
\Item{$P$}{pressure}
\Item{$R$}{gas constant}
\Item{$S$}{source term}
\Item{$T$}{temperature}

\end{list}

\subsection*{Subscripts}
\begin{list}{}{%
\settowidth{\labelwidth}{\textbf{$A,B,C,D$}}%
\setlength{\labelsep}{2. em}%
\setlength{\leftmargin}{\labelwidth}%
\addtolength{\leftmargin}{\labelsep}%
\setlength{\rightmargin}{0. cm}%
\setlength{\parsep}{\parskip}%
\setlength{\itemsep}{0. cm}\setlength{\topsep}{0. cm}%
\setlength{\partopsep}{0. cm}}

\Item{$abs$}{absorption into helium bubbles}
\Item{$i$}{hydrogen isotopes}
\Item{$EU$}{EU'97}
\Item{$F$}{fluid}
\Item{$G$}{gas phase}
\Item{$He$}{helium}
\Item{$HEN$}{heterogeneous nucleation}
\Item{$HON$}{homogeneous nucleation}
\Item{$LM$}{liquid metal}
\Item{$LLE$}{lithium lead eutectic alloy}
\Item{$M$}{membrane}
\Item{$S\!M$}{structural material}
\Item{$T$, $T_2$}{atomic, molecular tritium}

\end{list}

\subsection*{Superscripts}
\begin{list}{}{%
\settowidth{\labelwidth}{\textbf{$A,B,C,D$}}%
\setlength{\labelsep}{2. em}%
\setlength{\leftmargin}{\labelwidth}%
\addtolength{\leftmargin}{\labelsep}%
\setlength{\rightmargin}{0. cm}%
\setlength{\parsep}{\parskip}%
\setlength{\itemsep}{0. cm}\setlength{\topsep}{0. cm}%
\setlength{\partopsep}{0. cm}}

\Item{$b$}{bubble}
\Item{$0$}{pre-exponential}
\Item{$*$}{critical}

\end{list}

%%%%%%%%%%%%%%%%%%%%%%%%%%%%%%%%%%%%%%%%%%%%%%%%%%%%%%%%%%%%%%%%%%
%%% INTRO
%%%%%%%%%%%%%%%%%%%%%%%%%%%%%%%%%%%%%%%%%%%%%%%%%%%%%%%%%%%%%%%%%

\section{\label{sec:intro}Introduction}

Hydrogen isotope transport in matter is a critical issue in current nuclear technologies, from the point of view of design, operation and safety issues. For example, tritium inventory control and confinement is a key issue in nuclear fusion D--T reactors, concerning safety and the fuel cycle. Abundant literature can be found on hydrogen isotopes transport processes and on permeation barrier coatings, but there are no studies on the effect of nucleated bubbles at interfaces on hydrogen isotope permeation. The present work intends to give insight on impact of gas bubbles at interfaces on hydrogen permeation, that is bubbles as a permeation barrier.

\begin{figure}
\begin{center}
\includegraphics[angle=0,width=0.9\columnwidth]{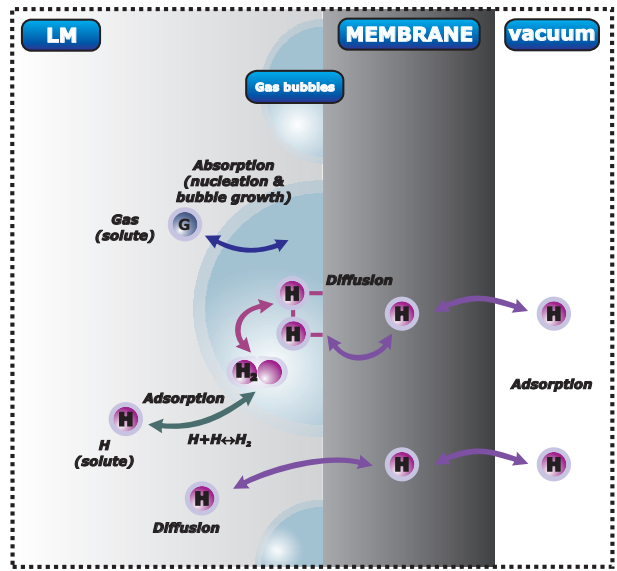}
\caption{Hydrogen isotope and helium transport phenomena in a permeation system.}
\label{fig:Phenomena}
\end{center}
\end{figure}

Helium nucleated bubbles at interfaces, e.g. at LLE -- structural material interfaces of a breeding blanket, can modify the permeation rates significantly. This work assesses this phenomenon, concluding that the impact on hydrogen isotope permeation is not negligible.

The model described across Sec.~\ref{Sec:models} has taken into account the following phenomena:

\begin{itemize}
\setlength{\itemsep}{1pt}
\setlength{\parskip}{0pt}
\setlength{\parsep}{0pt}
 \item{Helium nucleation at liquid metals--structural materials interfaces. Sec.~\ref{subSec:HEN}.}
 \item{Hydrogen isotopes permeation through liquid metals--structural materials interfaces in the presence of gas bubbles. Sec.~\ref{subSec:HDiff}}\\
\end{itemize}

The availability of a computational tool for tritium inventory evaluation within each sub-system of a fusion reactor, particularly in breeding units, is of great importance regarding design and operation. Key parameters affecting the fuel cycle design may be detected and, after experimental validation, models could be adjusted and T inventory quantified. The TMAP7 code has proven to be helpful for 0D and 1D simulations involving permeation processes and has been validated for tritium transport (see \citet{Longhurst08} and \citet{Ambrosek}). However, TMAP7 lacks the capabilities and flexibility of a CFD code.

An extensive review of hydrogen isotopes permeation barriers for metal structural materials (SM) in fusion power plants can be found in \citet{Hollenberg}. Measures of hydrogen permeation have been carried out in the ENEA research center at Brasimone, in the test sets Corelli~II \citep{Aiello01_1} and Vivaldi \citep{Aiello03,Aiello04}, where permeation barriers in contact with eutectic lead-lithium have been tested; experiments will continue in the TRIEX facility \citep{Aiello07}. \citet{Sedano1996} evaluated tritium permeation and extraction for the LIBRETTO-3 experiment (irradiation of LLE capsules coated with different permeation barriers). Tritium permeation barriers in contact with liquid LLE\ in stainless steel tubes was estudied by \citet{Forcey}, and \citet{Nakamichi} conducted several in-pile experiments on tritium permeation with ceramic coatings at the research reactor IGV. 1 M, in Kazakhstan, using liquid LLE\ alloys as the T source. 

In terms of theoretical modelling \citet{Fukada88} analytically modelled the permeation of hydrogen isotopes through a plate type metal window, for LM in
laminar flow. \citet{Farabolini} evaluated the main T flows in a fusion plant with HCLL breeding blankets, using a FEM code with a 2D simplified representation of the breeding unit to analyze the blanket system (see, as well, \citet{Gabriel}).
\citet{Gastaldi} developed a general model to analyse tritium release to the secondary circuit in a HCLL; FEM models were used to determine shape factors in order to correct for geometrical simplifications.

In the present work an specific model for He bubble nucleation at interfaces and bubble growth is presented, which has been implemented in the \OF CFD open source code (\citet{Jasak}) as a new solver. Implemented code is used to analyse He bubbles at interfaces effect on hydrogen isotope permeation.

In addition, in Sec.~\ref{Sec:Discus}, the following cases are exposed for analysis and raise conclusions related to current fusion technology designs, materials and operation conditions:

\begin{enumerate}
\setlength{\itemsep}{1pt}
\setlength{\parskip}{0pt}
\setlength{\parsep}{0pt}
 \item{Tritium permeation through Lithium Lead eutectic (LLE)--EU'97 interfaces with a homogeneous dispersed phase consisting in micro-bubbles at the interface Sec.~\ref{Sec:case1}.}
 \item{Hydrogen permeation sensitivity to a homogeneous dispersed phase consisting in micro-bubbles at LLE-stainless steels. Sec.~\ref{Sec:case2}.}\\
\end{enumerate}

\section{Implemented model in \OF}\label{Sec:models}

An adaptation of classic and well-known models to a CFD code have been implemented with the aim of predicting and analyse the effect of a gas phase at an interface in the form of micro bubbles on hydrogen isotope permeation. Hydrogen isotope absorption into the gas phase model, assuming both Diffusion Limited Regime (DLR), i.e. Sieverts' law, and Surface Limited Regime (SLR) was implemented in \citep{Fradera11} and it is applied in this work. Hydrogen isotope mass transfer between the gas phase and the solid structural material (SM) has been modelled following the same implementation as for the LM-gas phase. Mass transfer between LM and SM has been modelled following the law of solubilities. 

Gas bubble nucleation at interfaces has been modelled with the Self-Consistent Nucleation Theory (see \citep{Fradera11}). Bubbles are treated as a passive scalar, so, hydrogen isotope mass transfer processes have been calculated as a source term. All models have been coupled taking into account any interaction between the aforementioned phenomena.

\subsection{Surface Nucleation Model}\label{subSec:HEN}

\begin{figure}
\begin{center}
	\begin{subfigure}[b]{0.475\columnwidth}
			\includegraphics[angle=0,width=\textwidth]{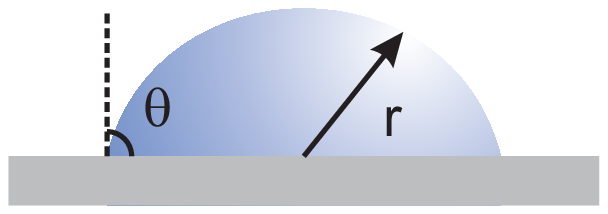}
			\caption{}
			\label{fig:HENFig}
	\end{subfigure}		
	\begin{subfigure}[b]{0.475\columnwidth}
			\includegraphics[angle=0,width=\textwidth]{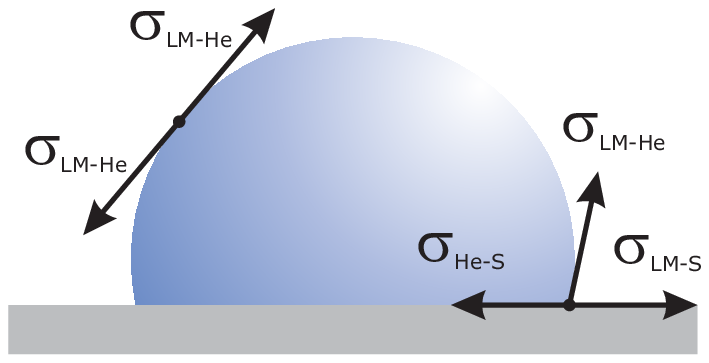}
			\caption{}
			\label{fig:surfTen}
	\end{subfigure}
	\caption{(a) A bubble at a flat surface on an active site. (b) Surface tension among phases for a spherical cap bubble on a flat surface.}
\end{center}
\end{figure}

In heterogeneous nucleation (HEN), bubbles are formed at preferential or active sites, like pores, walls (see Fig.~\ref{fig:HENFig}) or impurity particles under the necessary conditions. 

The Self-Consistent Nucleation Theory (SCT) \citep{Girshick90},  \cite{Girshick91} has gained acceptance due to its good results and simplicity.

The nucleation rate can be expressed as follows:

\begin{equation}
S_{SCT,HEN}\ =\ \dfrac{e^\Theta}{\psi}\, S_{CNT,HEN}\label{eq:SCTHEN}
\end{equation}
where $\psi$ is the supersaturation ratio and $\Theta$ the surface energy of one He atom in the cluster:
\begin{equation}
\Theta\ \equiv\ \dfrac{\sigma s_0}{k_{B}T}\label{eq:SCTcorr}
\end{equation}

and $S_{CNT,HEN}$ is the nucleation rate predicted by the CNT and formulated by \citet{Volmer}, \citet{Farkas}, \citet{Becker}, \citet{Zeldovich} and \citet{Frenkel} for homogeneous nucleation.

CNT treats the precipitates as spatially homogeneous so that the work of formation of a cluster of radius $r_{c}$ is given by the sum of the gain in free energy of the new stable phase and the cost in free energy due to the introduction of the interface (capillarity approximation). For HEN it takes the form:

\begin{equation}\label{eq:DGtot}
\Delta G_{tot}\, =\, \Delta G_{sur}\, +\, \Delta G_{vol}\, =f_{sur} 4\pi r_{c}^2\sigma_{G,LM} \, +\,f_{vol} \dfrac{4}{3}\pi r_{c}^3\Delta g_{vol}
\end{equation}

where $\Delta g_{vol}$ is the driving force for nucleation per unit volume of the new phase (i.e., it represents the free energy difference between the cluster and the dissolved states of one He atom divided by its volume). $f_{sur}$ and $f_{vol}$ are shape factors for the bubble surface and bubble volume depending on the type of active site upon which a bubble nucleates.

The driving force can be expressed by \citep{Frenkel,Wu,Gunton}: 

\begin{equation} \label{eq:Dgvol}
\Delta g_{vol} = \frac{- k_B T}{\upsilon_0} \ln(\psi)
\end{equation}

where $k_{B}$ is the Boltzmann constant, $T$ is the liquid metal bulk temperature and $\upsilon_0$ is the volume of one He atom in the cluster. $\psi$ is the supersaturation ratio, relating He concentration in LM ($C_{He,LM}$) with the saturation concentration $C^{sat}_{He,LM}$.

\begin{equation}
\psi=\frac{C_{He,LM}}{C^{sat}_{He,LM}}\label{eq:SatRatio}
\end{equation}

Maximum of eq.~\ref{eq:DGtot} with respect to $r_{c}$ is the barrier height to nucleation ($\Delta G^{\ast}$) eq.~\ref{eq:DGastHEN}, a magnitude strongly sensitive to $\upsilon_0$ and the surface tension ($\sigma_{G,LM}$). Classical theory assumes that surface tension is that of a planar liquid-gas interface, which is a gross overestimation that in some cases result in underpredictions of the nucleation rate. In the present work surface tension correction following the \citet{Tolman} model has been implemented to avoid the CNT overestimation. When a cluster surmounts the nucleation barrier it becomes stable and grows.

\begin{equation}
\Delta G^{\ast}_{HEN}=\dfrac{16\pi\sigma^{3}_{G,LM}}{3\, \Delta g_{vol}^{2}}f(\theta)\label{eq:DGastHEN}
\end{equation}

For spherical cap shape bubbles on a flat surface (Fig.~\ref{fig:HENFig}), the shape factor $f(\theta)$ (where $\theta$ is the wetting angle in Fig.~\ref{fig:HENFig}):

\begin{equation}
f(\theta)\ =\ \dfrac{1}{4}(2+\cos\theta)(1-\cos\theta)^{2}\label{eq:shapeFlat},
\end{equation}

Note that for spherical cap bubbles $f_{sur}\ =f_{vol}\ =f(\theta)$.

The number of atoms in the critical cluster reads:

\begin{equation}\label{eq:n-astHEN}
n^{\ast}_{HEN}\ =\ \frac{2\, \Delta G^{\ast}_{HEN}}{k_{B}T \ln(\psi)}.
\end{equation}

The expression for the nucleation rate for the CNT then takes the form:

\begin{equation}
S_{HEN}=S_{HEN}^{0}\, e^{-\Delta G^{\ast}_{HEN}/k_{B}T}\label{eq:rate-hen}
\end{equation}

where $S_{HEN}^{0}$ is the pre-exponential factor, proportional to the number of active sites.

The surface tension balance among the phases reduces the work of formation of new surface (see Fig.~\ref{fig:surfTen}). Hence, the energy barrier for HEN is lower than for Homogeneous Nucleation (HON), that is nucleation in the bulk liquid, ($\Delta G^{\ast}_{HEN}~<~\Delta G^{\ast}_{HON}$). Normally, HEN should take place at quite lower He concentrations than HON and, thus, be the preferred form of bubble formation.

Once one cluster reaches its critical size (given by the number of atoms $n^{\ast}$) it is assumed that it becomes instantaneously a bubble of radius $r_b^{\ast}$ and begins to grow.
It is also assumed that bubble size is so small that growth is controlled by diffusion and that inertial effects can be neglected. Thus, the rate at which He is added to or removed from a bubble $S_{He,b}$ is calculated as follows:
\begin{equation}
S_{He,b}=4\pi r^{2}_{b}\,f\, D_{He,LM}\,\biggl(\dfrac{\partial C_{He,LM}}{\partial r}\biggl)_{r=r_{b}}\label{eq:RateDiss2}
\end{equation}
where $D_{He,LM}$ is the diffusion coefficient. The concentration gradient $\left( \partial C_{He,L}/\partial r \right)_{r=r_{b}}$ is approximated to: 
\begin{equation}
\biggl(\dfrac{\partial C_{He,LM}}{\partial r}\biggl)_{r=r_{b}}\approx \frac{C_{He,LM}-C^{sat}_{He,LM}}{r_{b}}\label{eq:Cgrad}
\end{equation}
which is a simplification of the \citet{Epstein} model.

The whole process can be summarized as follows:

\begin{itemize}
\setlength{\itemsep}{1pt}
\setlength{\parskip}{0pt}
\setlength{\parsep}{0pt}
 \item{Helium solute reaches necessary supersaturation.}\label{P_1}
 \item{Helium solute begins to form unstable clusters at active sites.}\label{P_2}
 \item{Helium clusters reach the necessary size to be stable.}\label{P_3}
 \item{Helium stable clusters become a new gas phase and grow on the surfaces.}\label{P_4}\\
\end{itemize}

\subsection{Phenomena governing equations}\label{subSec:HDiff}

He governing equations are formulated for dissolved atomic He ($C_{He,LM}$), for He in the gas phase ($C_{He,G}$) and for the number of bubbles per unit surface ($N_b$):
\begin{equation}
\dfrac{\partial{C_{He,LM}}}{\partial{t}}\ =\  \nabla D_{He,LM}\nabla C_{He,LM}+ S_{He} - S_{He,abs}-S_{nuc}\label{eq:CHeLM}
\end{equation}
\begin{equation}
\dfrac{\partial{C_{He,G}}}{\partial{t}}\ =\  S_{He,abs}+ S_{nuc}\label{eq:CHeG}
\end{equation}
\begin{equation}
\dfrac{\partial{N_{b}}}{\partial{t}}\ =\ S_{nuc}\label{eq:Nb}
\end{equation}
where $S_{He,gen}$ is a source term taking into account He generation, e.g., by nuclear reactions, $S_{nuc}$ is a source term taking into account heterogeneous nucleation and $S_{He,abs}$ is the rate of He absorption due to the bubble growth mechanism. Note that all concentrations are referred to the LM volume so, e.g., $C_{He,G}$ is the He concentration in the gas phase per LM volume.  Note also that bubbles at surfaces are assumed to be static, thus, no convection term is included in Eq.~\ref{eq:Nb}.

Hydrogen isotopes $i$ and $i_2$ governing equations for each phase can be expressed as follows 
\begin{eqnarray}
\mbox{Liquid Phase:}& &\nonumber\\
\nonumber\\
\dfrac{\partial{C_{i,LM}}}{\partial{t}} &=& -(\textbf{u}\cdot\nabla C_{i,LM})+(D_{i,LM}\nabla^{2}C_{i,LM})\nonumber\\
& &+S_{i,gen}-S\!_{i,abs}\label{eq:CHLM}
\\
\dfrac{\partial{C_{i_2,G}}}{\partial{t}} &=& -(\textbf{u}\cdot\nabla C_{i_2,G})+\dfrac{1}{2}S\!_{i,abs}\label{eq:CH2G}
\\
\nonumber\\
\mbox{Solid Phase:}& &\nonumber\\
\nonumber\\
\dfrac{\partial{C_{i,S\!M}}}{\partial{t}} &=& D_{i,S\!M}\nabla^{2}C_{i,S\!M}\label{eq:CHS}
\end{eqnarray}
where subscript $G$ stands here for the whole gas phase and $S$ the solid phase or membrane. All concentrations are referred to the LM volume. $S_{i,gen}$ is the rate at which H is formed (by nuclear reactions).

Note that the interface condition linking eq.~\ref{eq:CHLM} and eq.~\ref{eq:CHS} is the law of solubilities (eq.~\ref{eq:solLaw}) and the continuity condition (eq.~\ref{eq:contEq}):

\begin{equation}
\dfrac{C_{i,LM \rightarrow S}}{k_{s,LM \rightarrow S}}\ =\  \dfrac{C_{i,S \rightarrow LM}}{k_{s,S \rightarrow LM}}\label{eq:solLaw}
\end{equation}
where $k_s$ is the Sieverts' coefficient.

\begin{equation}
D_{i,LM \rightarrow S}\dfrac{\partial{C_{i,LM \rightarrow  S}}}{\partial{n}}\ =\  D_{i,S \rightarrow  LM}\dfrac{\partial{C_{i,S \rightarrow  S}}}{\partial{n}}\label{eq:contEq}
\end{equation}
where $n$ denotes the normal coordinate to the surface. Note that concentrations are those at the interface LM-S.

\section{Analysis and Discussion}\label{Sec:Discus}

In this section, a sensitivity analysis to the fraction of area covered by He bubbles $a_{ratio}$ for a simple 1D system is exposed in order to show the effect of bubbles at the surface on the permeation process without the influence of convection. Moreover a 3D case showing such effect on a single pipe of a permeator is exposed for different stainless steels as structural materials. 

\subsection{One dimensional analysis of Hydrogen isotopes permeation with micro-bubbles at an interface}\label{Sec:case1}

\begin{figure}[ht!]
\begin{center}
\includegraphics[angle=0,width=0.5\columnwidth]{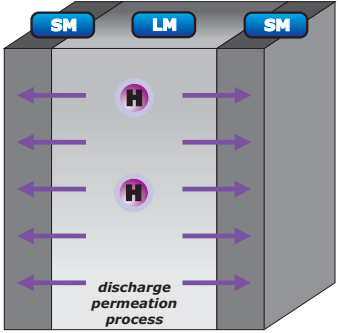}
\caption{One dimensional case configuration. Central LLE slab T discharge process through permeation process.}
\vspace{-10pt}
\label{fig:discharge}
\end{center}
\end{figure}

A simple case consisting of a LLE slab in contact with two EU'97 slabs as shown in Fig.~\ref{fig:discharge} has been chosen for the sensitivity analysis to the fraction of area covered by bubbles at a the LLE--EU'97 interface. At time zero the LLE slab is charged with a constant amount of tritium and EU'97 slabs are set to zero tritium concentration. T permeates through the EU'97 and a discharge happens until equilibrium between the LLE and the EU'97 is met. As an example, conditions have been set to those of a fusion reactor HCLL breeding blanket so as to show the effect under possible and probable realistic conditions.
LLE properties have been taken from \citet{MasdelesValls} at typical HCLL breeding blanket operating conditions (450$^o$C). Diffusion and Sievert's coefficients for tritium in LLE have been taken from \citet{Reiter}. Hydrogen absorption parameters in He have been taken from \citep{Terai90, Terai91} and \citep{Pisarev}, who modelled T release from molten LLE and compared results with \citet{Terai91} experimental data with well agreement. It must be noted that there is abundant literature on hydrogen isotopes transport parameters. However, transport coefficients show a wide span of values, specially for the solubility coefficient. Tritium diffusivity and solubility in EU'97 are taken from \citet{Esteban}.

T gradient at the EU'97 external boundary has been set to zero. This configuration also allows to verify mass conservation as the generated T stays within the system. At $t=0$~s T concentration is set to zero in the LLE and in the EU'97 slab. A constant T and He generation rate of 10$^{-7}$~mol/(m$^3$s),a typical value representing generation by nuclear reactions in a HCLL at nominal conditions, is set in the LLE domain.

\begin{figure}[ht!]
\begin{center}\vspace{-10pt}
\includegraphics[angle=0,width=1\columnwidth]{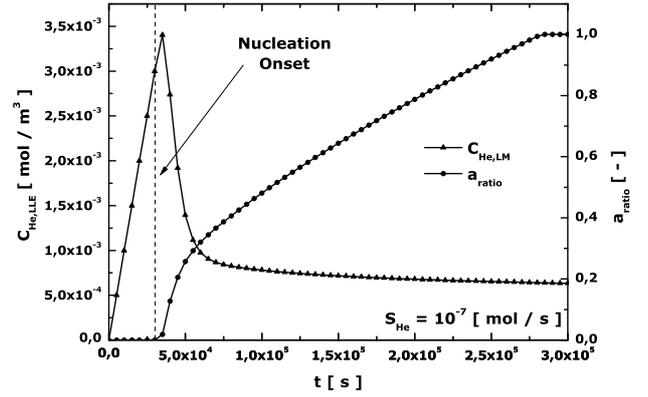}\vspace{-15pt}
\caption{$a_{ratio}$ and $C_{He,LLE}$ evolution at the LLE-EU'97 interface. Right axis is $a_{ratio}$.}
\vspace{-10pt}
\label{fig:aRatioDynamic}
\end{center}
\end{figure}

Fig.~\ref{fig:aRatioDynamic} shows the $C_{He,LLE}$ and $a_{ratio}$ evolution at the surface node. Nucleation begins after 3$\times$10$^{4}$s removing He from the LLE as more bubbles nucleate and former ones grow. No nucleation in the bulk LM ever occur as He is depleted below the necessary concentration for HON; only HEN at the interfaces occur. Concentration of bubbles evolution is not shown as it has the same tendency as $a_{ratio}$, reaching 2.15$\times$10$^{5}$~bubbles/m$^3$ at the end of the simulation. Void fraction at the boundary reaches a value of 8.6$\times$10$^{-4}$ Note that void fraction is very small even with the whole surface covered with bubble as it is referred to the volume of the control volume at the boundary.

\begin{figure}[ht!]
\begin{center}\vspace{-10pt}
\includegraphics[angle=0,width=1\columnwidth]{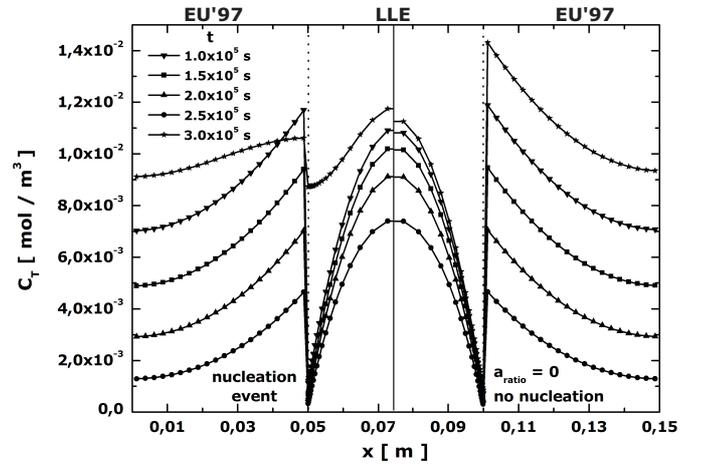}\vspace{-5pt}
\caption{T Concentration profiles comparison at different times. Right side: case without nucleation event. Left side: case with nucleation. (Nucleation event begins at t=7$\times$10$^{4}$s)}
\vspace{-10pt}
\label{fig:profiles_dynamicT}
\end{center}
\end{figure}
\begin{figure}[ht!]
\begin{center}\vspace{-10pt}
\includegraphics[angle=0,width=1\columnwidth]{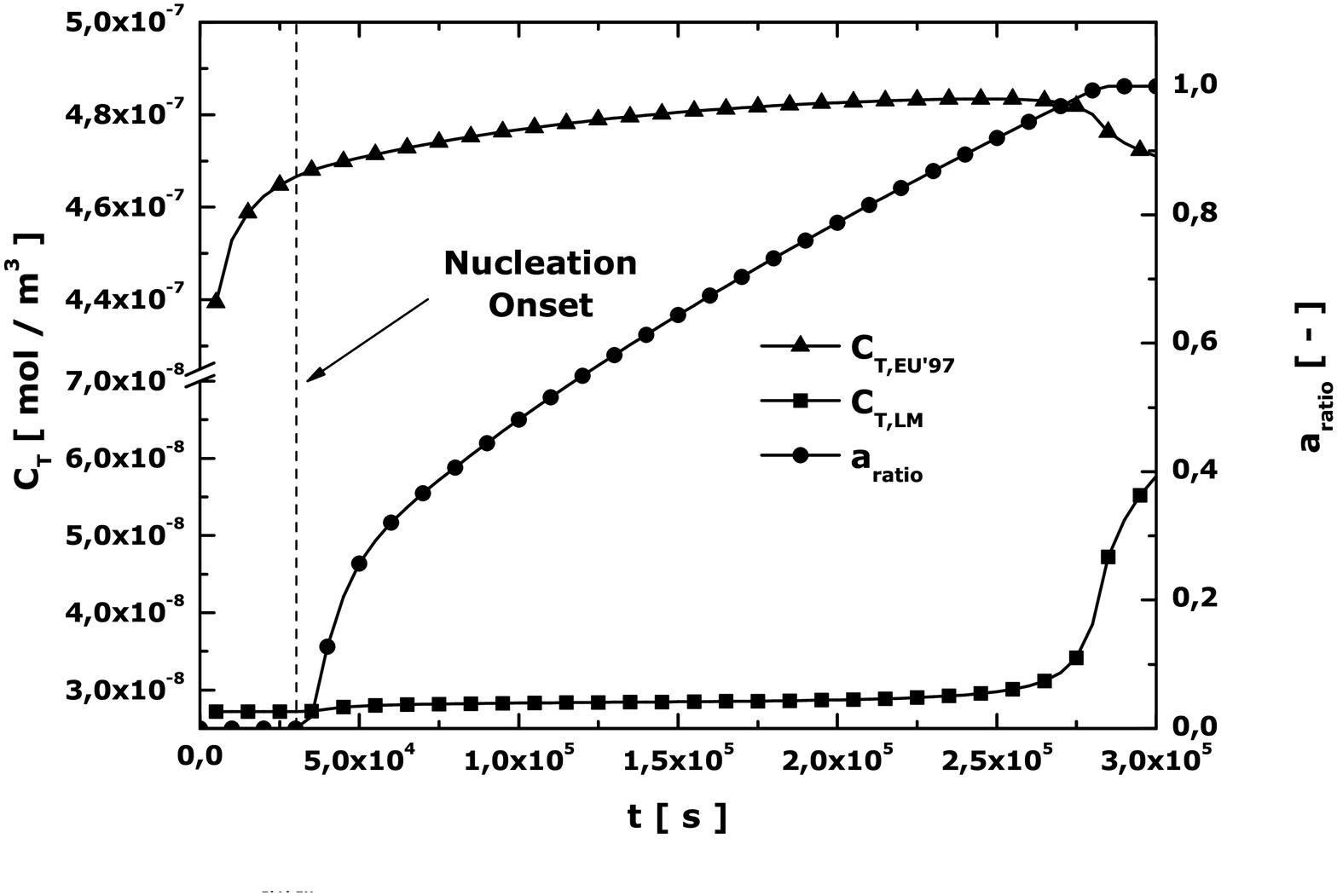}\vspace{-5pt}
\caption{T Concentration at the LM-EU'97 interface at different times. Right axis is $a_{ratio}$.}
\vspace{-10pt}
\label{fig:Trec_dynamic}
\end{center}
\end{figure}

Fig.~\ref{fig:profiles_dynamicT} shows the comparison between the transient case without nucleation (right of the figure) and that with a nucleation event (left of the figure). As $a_{ratio}$ increases, permeation is reduced, but the reduction can only be seen for times greater than 2$\times$10$^{5}$s; for $a_{ratio}=$1 permeation is reduced significantly (see also Fig.~\ref{fig:Trec_dynamic} tritium concentration at the interface). Concentration jump for times greater than 2$\times$10$^{5}$s show the effect of the bubbles on the permeation process. EU'97 no longer act as a by-pass for tritium because most of the surface is covered by bubbles. Thus, the He bubbles -- EU'97 mass transfer process is the controlling phenomenon. Despite the fact that T permeation is found to be reduced, as a nucleation event is also present, it is difficult to assess the permeation phenomenon itself.

A sensitivity analysis to $a_{ratio}$ for different and constant $a_{ratio}$ is exposed as a simplified case in order to show T permeation with more clarity. Bubble values, i.e. radius, bubble concentration etc, are taken from the previous case and set as constant values. He concentration is set to the saturation concentration and neither T nor He generation is set, so as to prevent bubble growth or re-dissolution. Hence, no nucleation event ever occur.

Initial T concentration in the LLE\ is set to 10$^{-6}$~mol/m$^3$ and to zero in the EU'97. It is expected that the recombination process will act as a very strong resistence for low T concentrations like the one set, but as long as there is a significant LLE--EU'97 interface, bubbles will not affect T permeation notably. Present simulations will show how T permeation is only significantly reduced for high $a_{ratio}$. System is let to evolve until equilibrium is met. 

He and T governing equations are simplified for this cases as follows:
\begin{equation}
\dfrac{\partial{C_{HeLLE}}}{\partial{t}}\ = D_{HeLLE}\nabla^{2}C_{HeLLE}\label{eq:CLsimp}
\end{equation}
\begin{equation}
\dfrac{\partial{C_{TLLE}}}{\partial{t}} = D_{TLLE}\nabla^{2}C_{TLLE}-S\!_{T,abs}
\end{equation}
\begin{equation}
\dfrac{\partial{C_{T_2,G}}}{\partial{t}} = \dfrac{1}{2}S\!_{T,abs}-\dfrac{1}{2}S_{T,ad}\label{eq:CT2Gsimp}
\end{equation}

\begin{figure}[ht!]
\begin{center}\vspace{-10pt}
\includegraphics[angle=0,width=1\columnwidth]{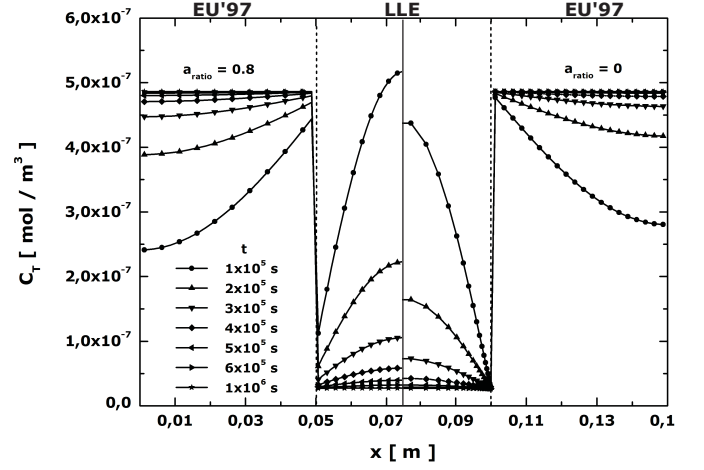}\vspace{-5pt}
\caption{Concentration profiles at different times for $a_{ratio}=$0.0 right and $a_{ratio}=$0.8 left.}
\vspace{-10pt}
\label{fig:aRatio}
\end{center}
\end{figure}

T concentration profiles at different run times for $a_{ratio}=$0 and $a_{ratio}=$0.8 assuming recombination limited, are shown in Fig.~\ref{fig:aRatio}. Note that simulations have been made for EU'97--LLE--EU'97 giving symmetric results. In Fig.~\ref{fig:aRatio} only half of the simulation results are shown for each $a_{ratio}$ case so as to compare results at a given time with clarity. 

For $a_{ratio}=$0 case, T is removed from the LLE\ and accumulates in the EU'97 until equilibrium is met. Process is mass conservative and at equilibrium system complies with the imposed law at the interface.

T concentration profiles at different run times for $a_{ratio}=$0.8 are shown in Fig.~\ref{fig:aRatio}. T is removed from the LLE\ and accumulates in the EU'97 until the same equilibrium as in Fig.~\ref{fig:aRatio} right side is met. Process complies with the imposed law at the interface. Comparing both cases, $a_{ratio}=$0 and $a_{ratio}=$0.8 it can be observed that bubbles at the interface slow down the permeation process significantly.
\begin{figure}[ht!]
\begin{center}\vspace{-10pt}
\includegraphics[angle=0,width=1\columnwidth]{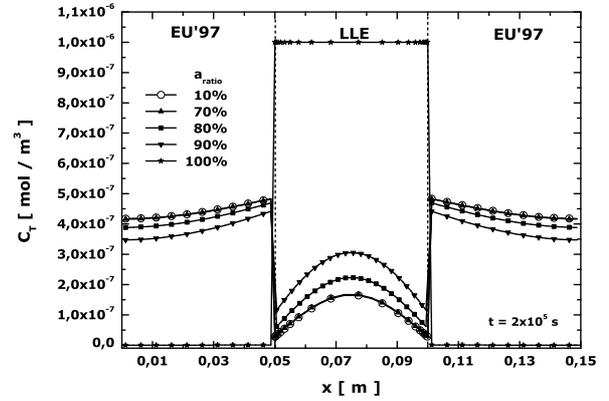}\vspace{-15pt}
\caption{Concentration profiles at different $a_{ratio}$ for $t=$2$\cdotp$10$^5$s.}
\vspace{-10pt}
\label{fig:aRatiot2e5}
\end{center}
\end{figure}

A sensitivity analysis to different $a_{ratio}$ is shown in Fig.~\ref{fig:aRatiot2e5}. The more the bubbles cover the interface the less T permeates to the EU'97 at a given time. Bubbles act as a permeation barrier. Note that mass is conserved for all simulated cases.

It must be noted that T concentration at the LLE-- EU'97 interface has been assumed to be always in equilibrium following the ratio of solubilities law. As a result, even with a recombination limiting process at the gas-bubble--SM interface, the permeation through the LLE-- EU'97 interface controls the mass transfer phenomenon. However, when $a_{ratio}$ reaches high values, up to 90\%, LLE-- EU'97 interface weight on the permeation is reduced rapidly. It can be concluded, under the simulated conditions, that mass tranfer is LLE-- EU'97 controlled unless most of the surface is covered with bubbles even for low T concentrations like the initial one set. For $a_{ratio}=$100\% the process is fully recombination limited and permeation is dramatically reduced. Fig.~\ref{fig:aRatiot2e5} shows the aforementioned effects.

\subsection{Hydrogen isotopes permeation through stainless steels pipes with micro-bubbles at an interface}\label{Sec:case2}

\begin{figure}[ht!]
\begin{center}
\includegraphics[angle=0,width=0.7\columnwidth]{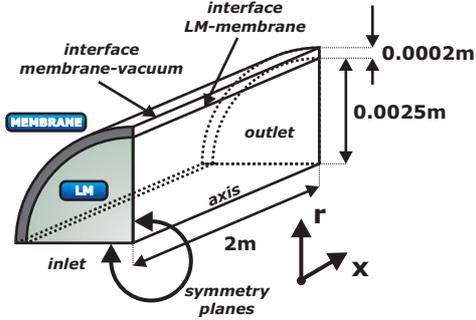}
\caption{3D pipe configuration for the pipe CFD simulation. Symmetry planes are used to save computational resources.}
\vspace{-10pt}
\label{fig:layout}
\end{center}
\end{figure}

Nucleation phenomenon at interfaces may not only be found in HCLL breeding blankets of a nuclear reactor, but also in hydrogen extraction systems and transport pipelines. Hence a single pipe as presented in Fig.~\ref{fig:layout}, with different stainless steels (EU'97, 304, 316 and MANET), has been chosen as a representative case. Hydrogen enters the pipe at a given and constant concentration of 10$^{-6}$~mol/(m$^3$s) and begins to permeate through the pipe walls. The concentration at time zero has been set to zero in all the LLE and EU'97 domains. 
Conditions have been set as for the one dimensional case in sec.~\ref{Sec:case1} for simplicity. Properties for 304 and 316 have been taken from \citep{Grant87,Grant88}, and for MANET from \citep{Serra}.

The present permeation system turns out to have an analytical solution in steady state if the external H concentrations is set to a constant zero value as follows,

\begin{equation}
C_{i,LLE}(x)=C_{i,LLE}(0)\ e^{-BL}\label{eq:CprofileLM}
\end{equation}
where $L$ is the total length of the pipe and $B$ reads,
\begin{equation}
B = \dfrac{2\ D_{i,S}\ k_{i,S}}{\textbf{u}\ k_{i,LLE}\ r_1\ ln(r_2/r_1)}\ \label{eq:BLM}
\end{equation}
where $r_1$is the inner radius and $r_2$ the external radius of the pipe and, $k_{i,F}$ and $k_{i,M}$ are the Sieverts' coefficients for the fluid and the membrane, respectively.

\begin{figure}[ht!]
\begin{center}\vspace{-10pt}
\includegraphics[angle=0,width=1\columnwidth]{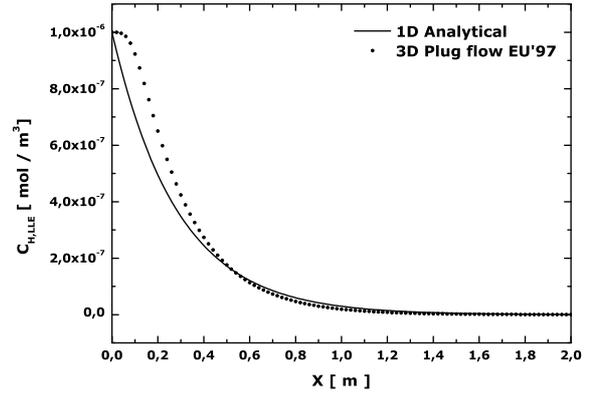}\vspace{-15pt}
\caption{Comparison between the analytical solution eq.~\ref{eq:solLaw} and the CFD solution for a 3D case for plug flow (Re=25) along the pipe and EU'97 as SM.}
\vspace{-10pt}
\label{fig:pComparison}
\end{center}
\end{figure}

In the present case both a plug flow profile and a fully developed laminar flow have been set along the pipe for Re=25, that ensures that LLE velocity is slow enough to achieve an efficiency close to 90$\%$ in the absence of bubbles at the interface. Fig.~\ref{fig:pComparison} shows the comparison between the analytical solution and this work's results for the 3D simulation with plug flow and EU'97 as SM. Agreement between results and the analytical solution is very good; the difference between both profiles lays on the fact that the simulation is 3D and, therefore, a H concentration profile across the pipe exists while in the analytical solution does not. 

\begin{figure}[ht!]
\begin{center}\vspace{-10pt}
\includegraphics[angle=0,width=1\columnwidth]{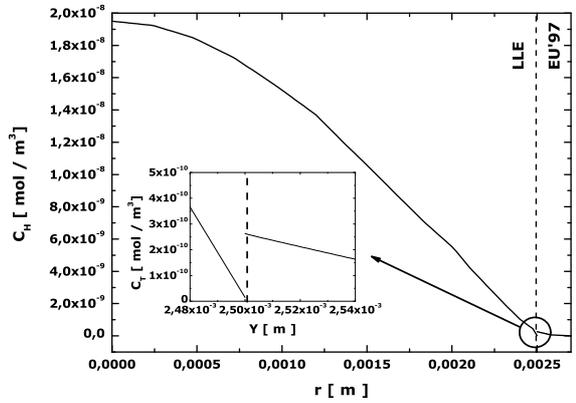}\vspace{-15pt}
\caption{T concentration profile across the pipe at X=1m (pipe center) showing the jump at the LM-EU'97 interface.}
\vspace{-10pt}
\label{fig:Yprofile}
\end{center}
\end{figure}

The H concentration profile across the pipe is shown in Fig.~\ref{fig:Yprofile}, showing the concentration jump at the interface LLE-EU'97. Concentration jump and profiles fully comply with the solubility law and the continuity condition.

\begin{figure*}
\begin{center}
	\begin{subfigure}[b]{0.475\textwidth}
		\includegraphics[angle=0,width=\textwidth]{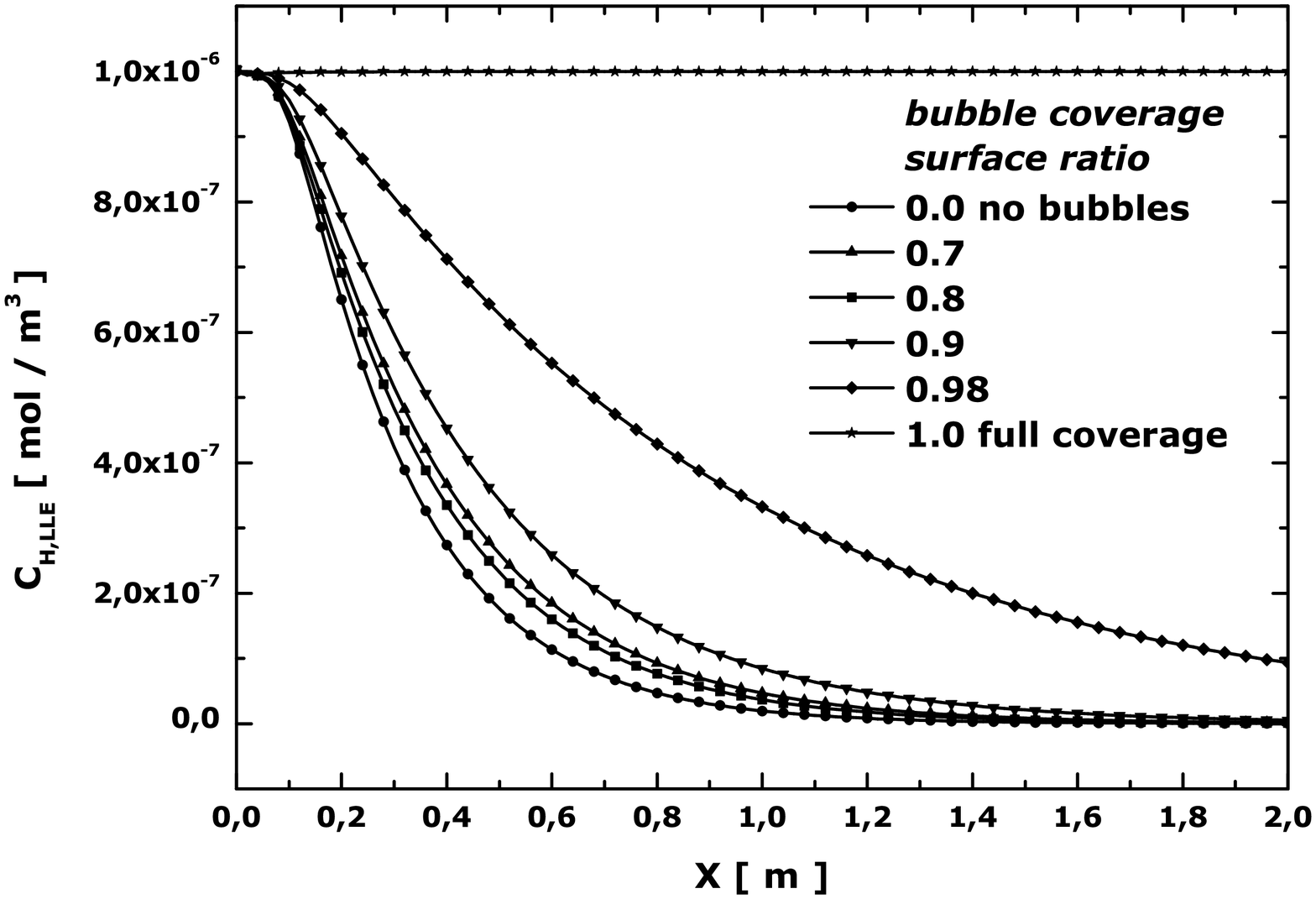}	
		\caption{}	
		\label{fig:CEU}
	\end{subfigure}	
	\begin{subfigure}[b]{0.475\textwidth}
		\includegraphics[angle=0,width=\textwidth]{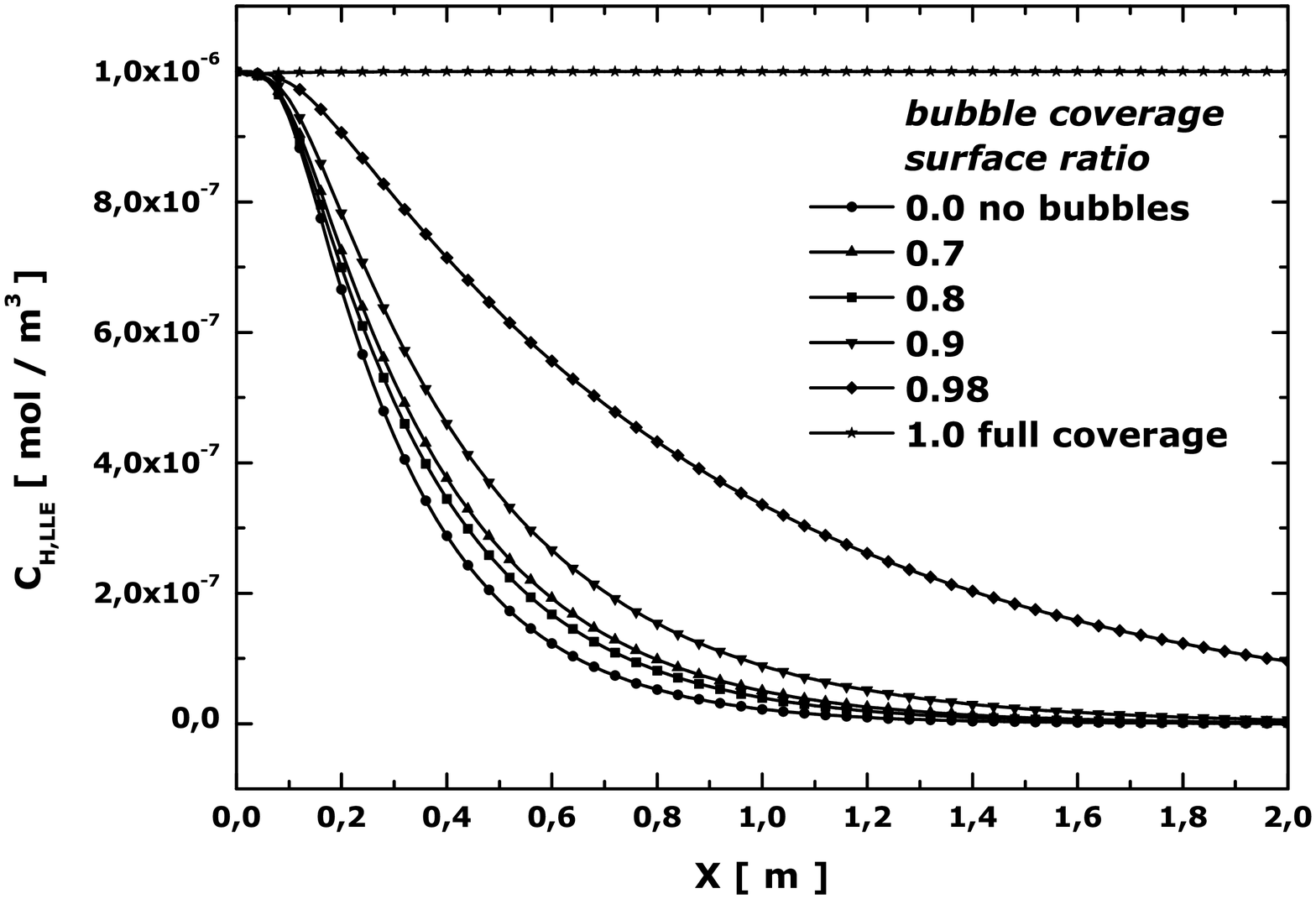}
		\caption{}
		\label{fig:C304}
	\end{subfigure}
	
	\begin{subfigure}[b]{0.475\textwidth}
		\includegraphics[angle=0,width=\textwidth]{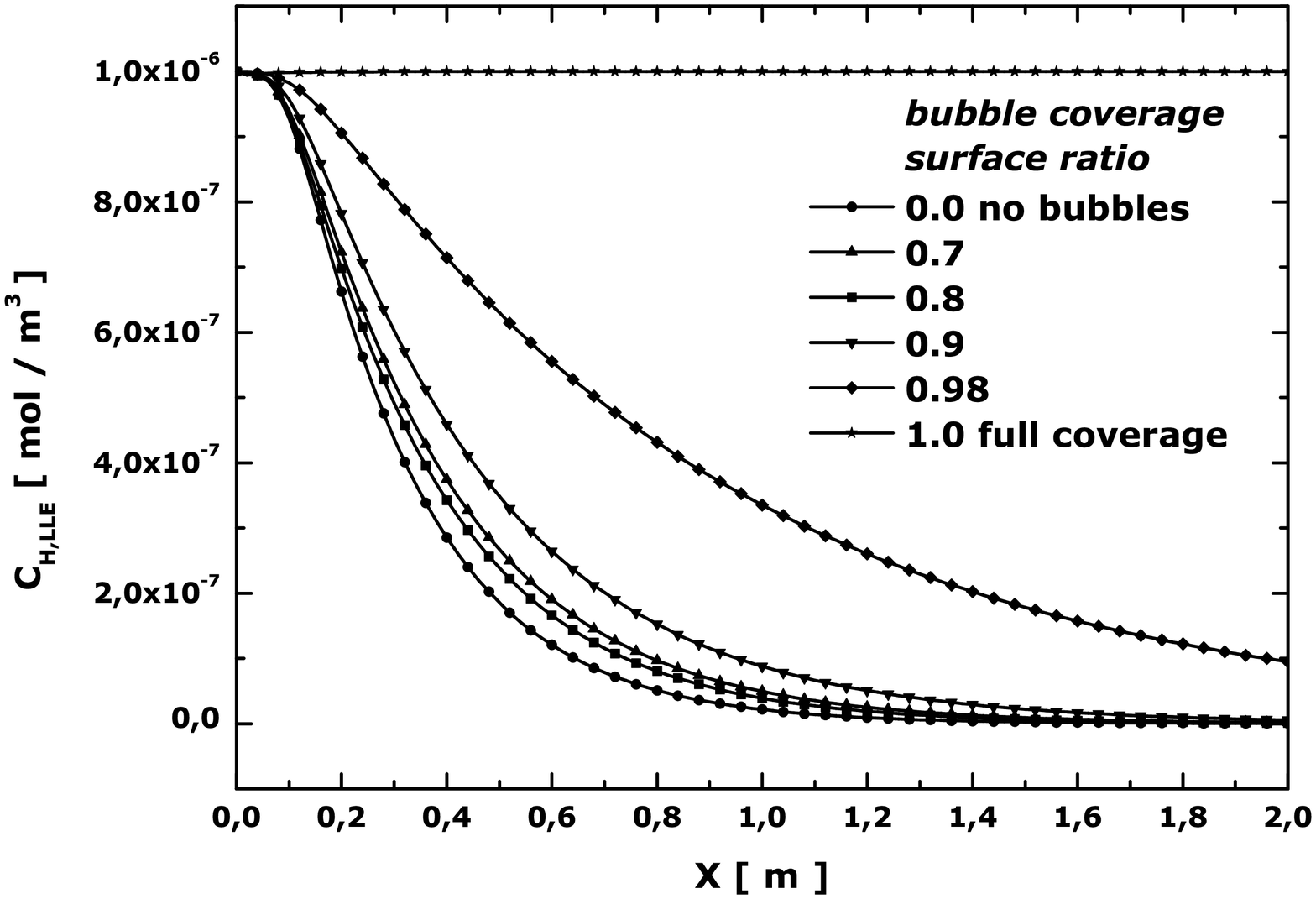}
		\caption{}
		\label{fig:C316}
	\end{subfigure}
	\begin{subfigure}[b]{0.475\textwidth}
		\includegraphics[angle=0,width=\textwidth]{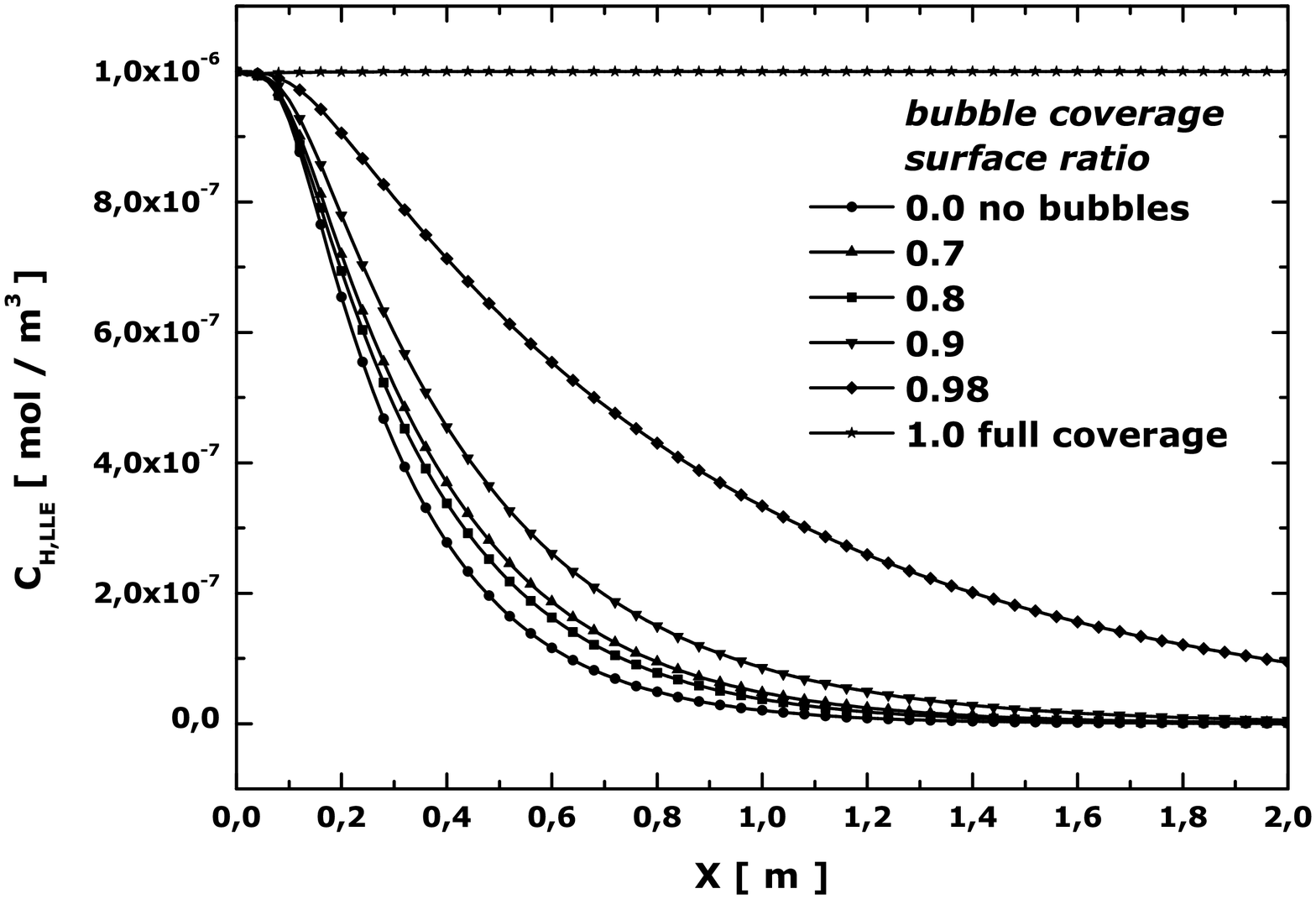}
		\caption{}
		\label{fig:CMANET}
	\end{subfigure}
	\caption{(a) Sensitivity analysis to $a_{ratio}$ for a fully developed flow (laminar flow Re=25) along the pipe and EU'97 as SM. (b) 304 as SM.(c) 316 as SM.(d) MANET as SM.}
\end{center}
\end{figure*}

A sensitivity analysis to different steels as SM for the pipe has been carried out showing well agreement and coherence between similar materials. Figs.~\ref{fig:CEU}, \ref{fig:C304}, \ref{fig:C316}, \ref{fig:CMANET} show the H concentration profiles for different $a_{ratio}$ and different steels. Note that the more the bubbles cover the interface, the smaller the amount of H that permeates becomes. At $a_{ratio}>0.9$ the percentage of H that has been removed becomes less that 90$\%$ and for a fully covered in bubbles interface, the permeation is almost inhibited.

\begin{figure}[ht!]
\begin{center}\vspace{-10pt}
\includegraphics[angle=0,width=1\columnwidth]{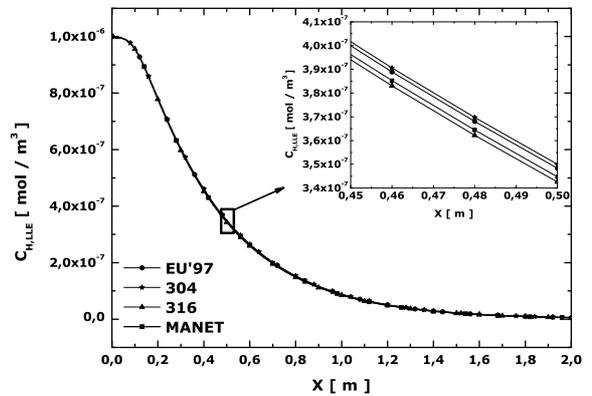}\vspace{-15pt}
\caption{Comparison between the different stainless steels for $a_{ratio}=0.9$. Zoom shows the small difference between materials regarding the effect on permeation.}
\vspace{-10pt}
\label{fig:mComparison}
\end{center}
\end{figure}

A comparison for the different steels for a pipe with the 90$\%$ of its interface covered in bubbles is shown in Fig.~\ref{fig:mComparison}. It must be noted that steels have similar Sieverts' coefficients and, thus, profiles are very similar with a deviation less than a 1$\%$.

\section{Conclusions}

%He nucleation at interfaces has been found to have a significant impact on hydrogen isotopes permeation through stainless steels for nucleation applications, specially for those involving LM like LLE. Nucleation control in LM for nuclear fusion and fission application may be of large importance regarding design, safety and operation. The present work demonstrates that He bubbles have an impact on hydrogen isotopes permeation and that more experiments supporting, verifying and validating present results are needed.

The presented work exposes a detailed, specific and coupled model transport phenomena implementation for transient He nucleation at interfaces, including self-consistent heterogeneous nucleation, curvature correction for He bubbles surface tension and tritium transport through interfaces. The resulting CFD code capabilities can be applied to many nuclear systems like those involving LLE and tritium (BBs), and may give valuable insight on this kind of system behavior through numerical experiments. 

He nucleation at interfaces has been found to have a significant impact on hydrogen isotopes permeation through stainless steels for nucleation applications, especially for those involving LM like LLE. Bubbles at interfaces act as a permeation barrier at high specific area ratios. However, at low specific area ratios, the surface free of bubbles behaves as a by-pass for permeation. Moreover, hydrogen isotopes concentration inside bubbles reaches saturation concentration quickly and, therefore, it can be assumed that permeation is limited to that concentration resulting driving force.

Nucleation control in LM for nuclear fusion and fission application may be of large importance regarding design, safety and operation. The present work demonstrates that He bubbles have an impact on hydrogen isotopes permeation and that more experiments supporting, verifying and validating present results are needed.

%% The Appendices part is started with the command \appendix;
%% appendix sections are then done as normal sections
\appendix

\section{Analytical solution for solute permeation through pipes} \label{App:App1}
The analytical solution for a solute like hydrogen, that permeates through a metallic or a homogeneous membrane can be derived by solving the following solute mass balance along the pipe in the carrier fluid or in the present work a LM:
\begin{equation}
\dfrac{\partial{C_{i,F}}}{\partial{t}}\ = -\textbf{u}\dfrac{\partial{C_{i,F}}}{\partial{x}}-S_p\label{eq:ODE_x}
\end{equation}

where $i$ denotes the solute, $F$ the fluid carrying the solute, $x$ is the coordinate along the pipe and $S_p$ is a source term taking into account the amount of solute that permeates through the pipe.
\begin{equation}
\dfrac{\partial{C_{i,M}}}{\partial{t}}\ = -D_{i,M}\nabla J_{i,M} = 0\label{eq:ODE_y}
\end{equation}

Applying a mass balance of solute across the membrane and assuming steady state (see eq.~\ref{eq:ODE_y}), an expression for the flux $J_{i,M}$ that permeates at a differential segment of the pipe $dx$ reads,
\begin{equation}
J_{i,M} = \dfrac{2\pi\ dx\ r_1\ D_{i,M}}{ln(r_2/r_1)}\ (C_{i,in}-C_{i,out})
\end{equation}
where $r_1$is the inner radius and $r_2$ the external radius of the pipe, $D_{i,M}$ is the solute's diffusion coefficient, $C_{i,in}$ is the solute concentration at the membrane's side of the interface and $C_{i,out}$ is the solute concentration at the membrane's external surface.

Assuming that all the solute that reaches the external surface is removed, i.e. $C_{i,out}=0$, and that the solubility law (eq.~\ref{eq:solLaw}) applies at the interface the source term for permeation (mol$_i$/m$^3$s) reads,
\begin{equation}
S_{p} = \dfrac{2\ D_{i,M}\ k_{i,M}\ C{i,F}}{k_{i,F}\ r_1\ ln(r_2/r_1)}\label{eq:Sp}
\end{equation}
where $k_{i,F}$ and $k_{i,M}$ are the Sieverts' coefficients for the fluid and the membrane, respectively.

Hence, substituting eq.~\ref{eq:Sp} in eq.~\ref{eq:ODE_x} and integrating along the pipe the following expression for the solute profile along the pipe is found:
\begin{equation}
C_{i,F}(x)=C_{i,F}(0)\ e^{-BL}\label{eq:Cprofile}
\end{equation}
where $L$ is the total length of the pipe and $B$ reads,
\begin{equation}
B = \dfrac{2\ D_{i,M}\ k_{i,M}}{\textbf{u}\ k_{i,F}\ r_1\ ln(r_2/r_1)}\ \label{eq:B}
\end{equation}
%

%% References
%%
%% Following citation commands can be used in the body text:
%% Usage of \cite is as follows:
%%   \cite{key}          ==>>  [#]
%%   \cite[chap. 2]{key} ==>>  [#, chap. 2]
%%   \citet{key}         ==>>  Author [#]

%% References with bibTeX database:

\bibliographystyle{model1a-num-names}
%\bibliography{<your-bib-database>}

%% Authors are advised to submit their bibtex database files. They are
%% requested to list a bibtex style file in the manuscript if they do
%% not want to use model1a-num-names.bst.

%% References without bibTeX database:

\end{document}